# Simulation of Blockchain based Power Trading with Solar Power Prediction in Prosumer Consortium Model

Kaung Si Thu, Weerakorn Ongsakul
Department of Energy, Environment and Climate Change
School of Environment Resources and Development (SERD)
Asian Institute of Technology
Bangkok, Thailand
a.kaungsithu@outlook.com, ongsakul@ait.asia

*Abstract*—Prosumer consortium energy transactive models can be one of the solutions for energy costs, increasing performance and for providing reliable electricity utilizing distributed power generation, to a local group or community, like a university. This research study demonstrates the simulation of blockchain based power trading, supplemented by the solar power prediction using MLFF neural network training in two prosumer nodes. This study can be the initial step in the implementation of a power trading market model based on a decentralized blockchain system, with distributed generations in a university grid system. This system can balance the electricity demand and supply within the institute, enable secure and rapid transactions, and the local market system can be reinforced by forecasting solar generation. The performance of the MLFF training can predict almost 90% accuracy of the model as short term ahead forecasting. Because of it, the prosumer bodies can complete the decision making before trading to their benefit.

## I. Introduction

Considering of rising demand for electricity and greenhouse gas (GHG) emissions, countries around the world are setting objectives for reducing GHG emissions, enhancing electricity quality and increasing renewable energy production. Consequently, distributed generation (DG) technologies have been extended that will provide effective renewable electricity across local power grids [1], however, a sufficient impact of penetration of renewable energy will be the primary energy supply. The market could be volatile frequently and so as the different energy sources. Community grids are being discussed in the literature and are being used for implementing the current power generation and the trading system currently. Asian Institute of Technology (AIT) in Thailand has committed and operational captive generation units and having the peak loads during the daytime with the academic and administrative sections accounting for the consumption, it appears as a good customer for solar-based power production. The residential systems, if equipped with solar power, can generate power during the day and supply the same to the other sections of AIT or the power grid and get compensated with electricity at night, providing opportunities at incentives as well as avoiding the need for storage. Numerous such options can be evaluated if AIT can be viewed as a community level micro or macro-grid, feasible provided a proper market model featuring consumers, producers and prosumers are employed and a proper trading mechanism is implemented. The idea of a community grid is well associated with existing energy policies. A grid facilitates the integration of energy storage systems (ESS) and renewable energy sources (RES) at the usage level, intending to increase the quality, reliability and performance. AIT can be considered as local grid which is especially campus microgrid or community grid. The AIT Grid is conventionally connected with 22 kVA grid from Pathum Thani Provincial Electricity Authority (PEA). The radical type of electrical power distribution system is applied in the campus distribution system. Generally, it is comprised of 15 substations and 92 main nodes, which does not include the several quantities of loads under each node according to the last updated period. The electrical power distribution system at AIT is integrated with DERs, which are rooftop solar power generation systems, generator and EV stations. This research study intended to focus on a blockchain, and AI based power trading in prosumer consortium model. The simulation of prosumer consortium model from the existing power distribution system of Asian Institute of Technology had elaborated with the scheme of blockchain-based power trading body with the demonstration of the trading based on the actual data of the prosumer bodies. An artificial neural network which is Multi-layer Feed Forward (MLFF) was performed to forecast the solar PV power generation of prosumer bodies by one day ahead, moreover, short-term prediction of solar generation on the day of the energy trading for both bodies was applied.

## A. Prosumer Model and Blockchain

Conventional electricity operation system becomes unsuited with the progression market behaviour. Therefore, new market models are now conducted not only in research but in practice with the competitive market. In which, the market can be generally distinguished into (1) centralized electricity market and (2) decentralized electricity market. The power trading deal may be intraday or future ahead trading focused on the opening of the market, anonymity in asset valuation and no credit risk at all. Consortium blockchains, on the other hand, are distributed on various hardware operated by separate owners in a decentralized way.

The energy market has undergone major reforms to globalization and innovation in the past decade to increase economic performance. These developments have culminated in the introduction of a bulk power system in many countries. The complexity of a community grid depends on the set of small independent business players and a partnership that involves all of the supply-side and market agencies. Its operation depends either on the end consumer, supplier, DSO and generation sources or the operator of independent power producer. As for now, the market of microgrid or community grid can be generally distinguished into prosumer consortium, free market and monopoly market.

Prosumer consortium market model is concerned with either environment where there is sufficient renewable sources or surge electricity market. Single or multiple consumers will operate the microsource to decrease the energy price and increase the profit of the sales within the community. DSO actively controls microgrid operations only by implementing criteria and charges on the microsource owners. The size of renewable generations and storage system is usually smaller in a prosumer consortium grid system and dispersed. Market cash flow in the prosumer consortium model is mentioned in Fig 1. It indicates the trading of electricity within the grid system and the wholesale market.

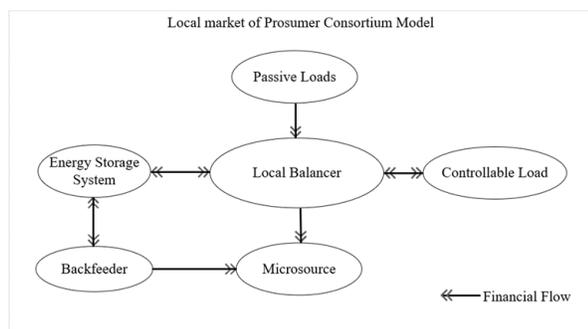

Figure 1. Local financial flow of prosumer consortium market model.

Community grid has a decentralized power network that consists of integrated green and conventional energy sources, and storage in smart buildings with energy management systems. Though the local consumers are connected with main electricity grid, it has the ability to produce and use their power generation. The circumstances allow the development of integrated DER system to the community grid users from commercial, residential to institutes. According to Basak et al. [2] onsite electrical production can reduce the electricity price by 25% at most which gives reliable service and reduce the line losses to 7%. However, adaptation and development have to be made on the switching mode, infrastructure components, quality control and protection [3]. Prosumer model and liberalized market, rather than DSO monopolies, need more regulatory and financial support.

Information Technology and Communication (ICT) is an essential part of future control systems. Without any doubt, complex information technology and integrated communication networks need to enable the management and maintenance of future power grids. Successful grid management should be focused on current connectivity networks to reduce costs. The new trend in designing information systems is service-oriented architectures. Web service is described as a software framework designed to facilitate open source machine-to-machine communication system [4].

Furthermore, information between consortium nodes is not inherently homogeneous, as some blockchains require private transactions relating to information divergence. Data ledger structure is also the private or collective cooperative, a blockchain's database refers to a connected collection of blocks representing transactions often known as the ledger system. A hash value is the primary component of the validity of the blockchain and is determined by one-way hash function that links arbitrary size data to a fixed size non-invertible hash value. Fig 2 presents distributed ledger structure, blockchain is one component of a decentralized network that holds shared records. The database includes block records, instead of merging them into a single register. In sequential, each new block is added to the preceding block using a hash algorithm which effectively makes it impossible to modify the details. This mechanism requires blockchains to be used as ledgers and someone with the necessary permissions will access and confirm. It may spread these decentralized ledgers across several locations, governments, or organizations.

A blockchain can submit data to each other directly that links the users and along with their data structure together. Members join the network through their application node to blockchain. A node checks the validity of the blocks by evaluating all of the hashes while accessing the database for the first instance and continues checking each new block. Cryptographic asymmetric key pair determines the identity of a member. This extracts the cryptographic key to access the unique address, which serves as the user identifier. The private key has been used to sign documents and to ensure their validity in which other members must use the valid public key to validate the signature. A node hereby submits a request to the network for a transaction to link data to the database. For any implementation, a transaction's primary data fields are the recipient and receiver addresses, the data values being exchanged, and the sender's signature. Then, the requests for the trades are provided by multiple different nodes named miners, who are often referred to on consortium blockchains as block generators or validators.

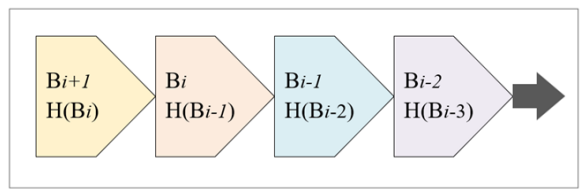

Figure 2. Simplified linked block with hash position.

It must recognize and resolve the key motivating concerns and possibilities within the energy sector of blockchain technology mistreatment and disseminate these approaches around the market in order to have a significant and enduring financial and social impact. For instance, a distributed database system could be used to verify the data, the source of energy consumption, thereby ensuring that it is safe for the environment. It can also be used to exchange electricity from LO3 Electricity in Brooklyn at electricity grid level, between local suppliers and customers. More savings can be conceived in the household, where appliances can schedule their energy consumption to minimize costs and exchange data efficiently between them [5].

A study, Oh et al., presents the deployment of a fully decentralized, blockchain based powered energy trading network across the establishment of consumer, seller and server nodes as well as database [6]. The program allows purchaser / seller sales data to consumers and promotes the transaction of power through smart contracts. That is to note, the basic operating model of the console is designed to optimize the process for the transaction. Then authorized transaction records are stated in and shared in a public ledger. Within a blockchain, multichain will identify and use two or three properties, so it may be power and money trading.

### B. Solar PV Generation and Prediction by AI

AIT has the potential for photovoltaic power generation which will help balance the load and reduce the electricity bill anyhow. According to Global Solar Atlas, power generation can reach to 4.018 kWh/kWp per day. Additionally, two prosumer buildings with rooftop solar panels, AIT Library building and AIT Energy building, were only used to implement in this research study though the campus has other renewable energy producer and prosumer. The production by the energy building and the library were 87,62.63 kWh 64,550.7 kWh individually in 2019. The impact of artificial intelligence on the energy paradigm of the prosumer consortium has had a significant effect on the decentralized structure, rather than centralized control, where solar energy production forecasting, load demand forecasting, and electricity price forecasting. It leads the day ahead or an hour ahead to estimate demand and power supply, increasing production prices and as well as the lowest point to the highest level of energy bill [7].

Santofimia-Romero et al. [8] states that the main feature of the smart grid is the need for actual time sharing of information, which will be focused on some kind of connectivity network that can facilitate the convergence of distributed and diverse systems. Artificial Neural Network (ANN) consists of simulation of biological neural network activities. In this way, a neural network contains a set of entangled nodes that are impartial processing units that have two weights, an input as well as a weight associated with it. Most recognizable aspect of the neural networks is that they can be taught to recognize certain patterns of data, rather than being programmed to perform certain tasks. The analysis has shown that ANN provides the conventional regression method with a successful outcome. Due to the extremely non-linear nature of the relationship between radiation and other meteorological data, ANN can be a reasonable prediction method [9]. Significant algorithmic change that greatly improved the feedforward network performance was the replacement of concealed sigmoid function with piecewise linear units, such as rectified linear units [10].

## II. METHODOLOGY

### A. Prosumer Models and Market Mechanism

The energy market is simulated in MATLAB Simulink based on the real data of solar power generation and consumption of AIT Library and AIT Energy building. The power output of solar generation can be simulated randomly, however, in this study, real data is applied to simulate the solar power generation. Generally, as in Fig 4, the surplus power produced by solar generation of AIT Library nodes is not expected to satisfy to trade, However, the surplus power generation at AIT Energy building is around 5 kW maximum which is enough to make power trading to AIT Library building as illustrated in Fig 3. The installed capacity of solar generation at AIT Library building is 60 kW, and solar generation capacity at AIT Energy building was 4.2 kW before October 2019 and now 12.0 kW after increased in capacity.

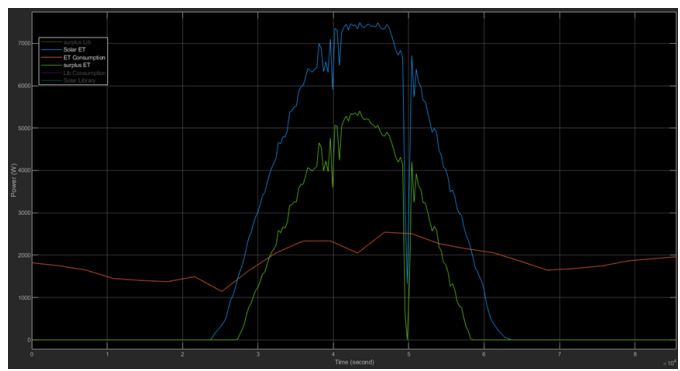

Figure 3. Solar power generation (blue), consumption (orange) and surplus power (green) of AIT Energy building.

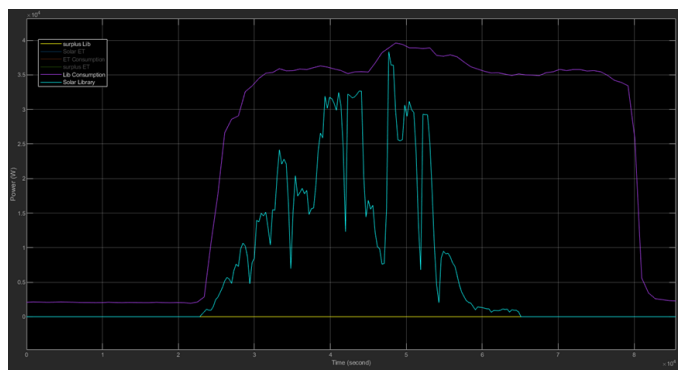

Figure 4. Solar power generation (cyan), consumption (violet) and surplus power (yellow) of AIT Library building.

The simulation contains imported generation and consumption from the actual data. Each node has minor controllers and one major controller. Solar generation data and consumption data of both prosumer nodes are imported from the actual data. Algorithm 1 is implemented at the solar power generation and prosumer's consumption, which can perform the detection of surplus power left after the consumption of

```
Algorithm 1: Detecting surplus power
Result: Surplus power
while simulation do
    if demand = grid then
        Pgrid = grid;
        Genleft = Pgen-(demand - Pgrid);
    else
        if demand > grid then
            Pgrid = grid;
            Genleft = Pgen - (demand - Pgrid);
        end
    if Genleft ≤ 0 then
        Genleft = 0;
        surplus = Genleft;
    end
end
```

```
Algorithm 2: Energy deduction and Transfer
Result: Transfer energy, deducted energy
while simulation do
    if P_ET ≤ 0 then
        DemandbyET = 0;
        Surplus_Lib = P_Lib - DemandbyET;
    else
        if P_ET >0 then
            DemandbyET = ET_demand;
            Surplus_Lib = P_Lib - ET_demand;
    end
    if P_Lib ≤ 0 then
        DemandbyLib = 0;
        Surplus_ET = P_ET - DemandbyLib;
    else
        if P_Lib >0 then
            DemandbyLib = Lib_demand;
            Surplus_ET = P_ET - Lib_demand;
    end
end
```

where,
| | |
|---|---|
| *demand* | Power demand by prosumer itself. |
| *Pgrid / grid* | Feed in power by main grid. |
| *Pgen* | Power generation from solar PV. |
| *Genleft / surplus* | Surplus power after the consumption of prosumer itself. |
| *P_ET* | Surplus power input by Energy building (AIT ET). |
| *P_Lib* | Surplus power input by Library building (AIT Lib). |
| *DemandbyET* | Demand data output by user input. |
| *DemandbyLib* | Demand data output by user input. |
| *Surplus_Lib* | Surplus power after demand input. |
| *Surplus_ET* | Surplus power after demand input. |
| *ET_demand* | Demand data input by Energy building. |
| *Lib_demand* | Demand data input by Library building. |

prosumer itself. Algorithm 2 performs the energy transfer and deduction of energy from surplus energy that was detected by algorithm 1.

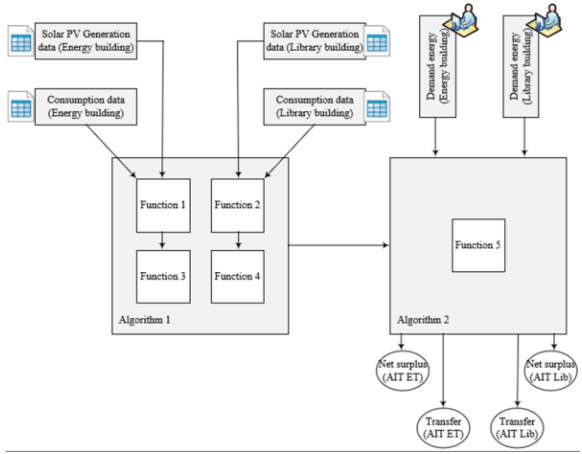

Figure 5. Diagram of MATLAB simulation structure for energy trading.

Architecture of energy trading simulation in MATLAB can be stated in Fig 5. As it is mentioned in earlier, all the performing functions are embedded in the minor (Function 1, 2, 3 and 4) and the major control (Function 5), and its final generation and the estimated consumption data will be performed as in algorithm 1 and the output signal will be continued to algorithm 2 to start a transaction. Then deducted energy by buyer and remaining energy will then be displayed to nodes as output again.

B. *Blockchain Mechanism in the System*

Blockchain system of this study is shaped in MATLAB environment with basic and simple method. The blockchain application in the MATLAB is implemented in each prosumer node of the system, therefore, a single computer is used to create the blockchain transition of both nodes effectively. The overview concept of the blockchain connection in AIT library and AIT energy building is explained in Fig 6. The function code that follows advanced contract is being deployed in order to trade by using token per unit.

Blockchain connection is established between AIT Library and Energy Building as localhost connection. Both nodes have each blockchain application and displayed command window. Blockchain application is created in MATLAB with graphical user interface (GUI). The callback functions in GUI for performing the transaction is embedded in MATLAB ".m" file. The general blockchain framework is the reference from MATLAB community. Layout and command in GUI of prosumer nodes are shown in Fig 7. GUI presents MATLAB code to function the blockchain transaction and token transfer. Each GUI is separated as different command window and MATLAB.

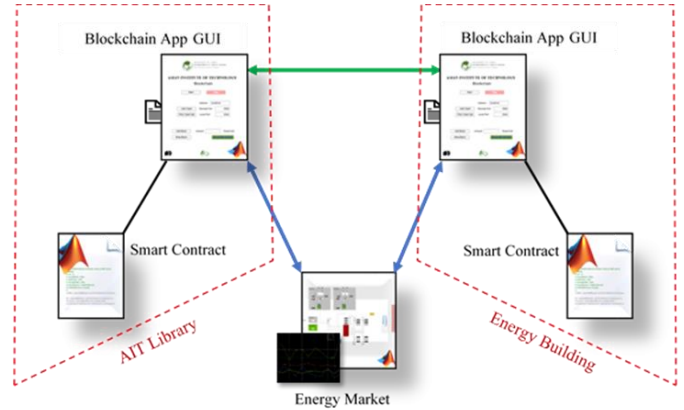

Figure 6. Overview linking of two prosumer nodes with blockchain application and energy market.

An advanced contract is implemented and adapted by smart contract from ERC-20 Token Standard in Ethereum improvement. However, none of cryptocurrencies were used to trade in the system, and it will be used as simple digital token system which follows the regulation of the advanced contract. This allows us to transfer and approve tokens wallets to decentralized exchanges.

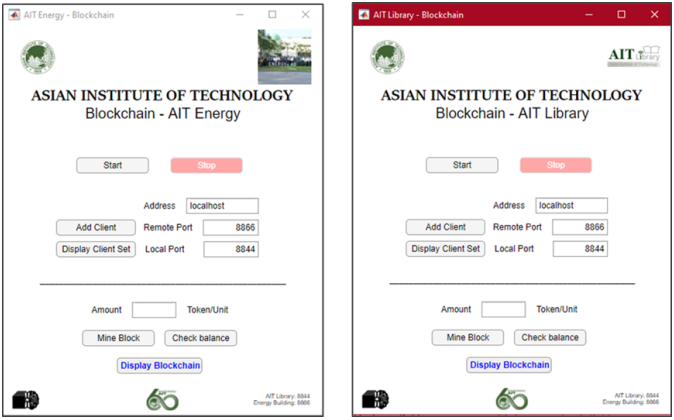

Figure 7. Layout of blockchain GUI in two prosumer nodes - AIT Energy building (left) and AIT Library building (right).

Since a single computer is used as mentioned, it is required to start two separate MATALB software. MATLAB function code and files for both prosumer nodes are kept in different directories as well as advanced contract and virtual wallet. The following processes allow completing nodes establishment and transaction between two prosumer nodes –

- Initiate the process of the blockchain application at AIT library and Energy building individually during the run of simulation of energy market at different MATLAB software.

- Localhost connection is used to connect local port and remote port and they were assigned to AIT library and Energy building respectively.

- Perform Add Client after the selection is made. The confirmation will appear in command windows of both nodes.

- Token balance must be check before making any transaction. Only with enough balance can start the transaction, otherwise, the process will be terminated.

- Meanwhile the worth of each power unit is one token, the required unit can be requested in the GUI. Sufficient surplus power can complete the transaction. After that the mining process will be initiated.

- After the mining at a node, messages will be announced in both nodes. This allows the transfer of requested power unit from seller to buyer. Then the transmitted block, can be view from command windows of both prosumer nodes. Both nodes can send the token to and from.

Each block creation takes few seconds to issue index, timestamp of block creation, transfer credit data, nonce, own hash string, and previous hash string. The first block is called "*The origin*" and it is the very initial creation of blockchain system. Current hash value is generated based on the combination of index values to previous hash value. The mining process will start once both nodes are sufficient to trade and make agreement. Tokens from buyer will send to the seller balance as soon as the block is successfully created and announce to every connected node. The requested power will transfer to the buyer network after the seller receives the respected tokens then it will make a complete energy transition.

### C. Data structure and Neural Network Training

Multi-layer feedforward neural network (MLFF) has neurons which are grouped into layers to form multiple layers, in much each layer has connection. It generally consists of three types of layers -the input layer, the output layer and the hidden layers. One of the approaches to use as activation function is the rectified linear activation function (ReLU) which rectified for hidden layers. ReLU implementation can reasonably be considered one of the benchmarks in the deep learning. The rectified linear activation function is a simple equation which simply returns the given input as input [11]. Based on the inputs, every synapse has its own weight which determines the degree of performance of a neuron [12].

The rectified linear activation function of the neurons inside hidden layers is used in (1). The notation of $x$ defines the input of $i^{th}$ unit, the value of weight - $w$ in $i^{th}$ and $j^{th}$ unit. The value after a hidden layer can be stated as $u$. Subsequently, the values of after activation function with hidden layers can be denoted in (2). Besides $\theta$ shows the bias value. The output layer can be seen in (2), where $y$ is output value with respect to $k^{th}$ unit, and the architecture of the neural network can be seen in Fig 8.

$$\emptyset(x) = \max\{x, 0\}$$

$$u_j = \sum_{i=1}^{n} x_i\, w_{ij} \qquad (1)$$

$$h_j = \emptyset(u_j + \theta_i)$$

$$h_j = \emptyset\left(\sum_{i=1}^{n} x_i\, w_{ij} + \theta_i\right) \qquad (2)$$

$$y_k = \emptyset\left(\sum_{j=1}^{m} u_{jk}\, h_j + \theta_k\right)$$

$$y_k = \sum_{i=1}^{m}\left(\frac{1}{1 + e^{(-\sum_{i=1}^{n} x_i\, w_{ij} + \theta_i)}}\, u_{jk}\right) + \theta_k \qquad (3)$$

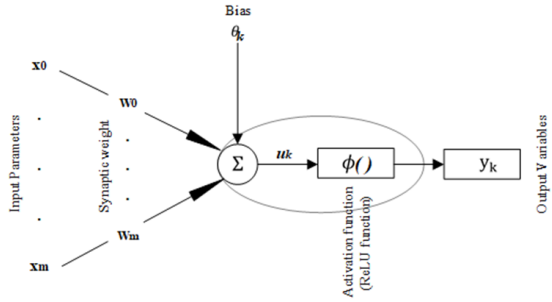

Figure 8. Architecture of MLFF Neural Network.

Input parameters which will be used to train the model such as solar irradiance, air temperature, relative humidity, and rainfall can be defined as –

- $x_1$ - Solar irradiance (W/m$^2$)
- $x_2$ - Average air temperature (°C)
- $x_3$ - Average relative humidity (%)
- $x_4$ - Average rainfall (mm)

Prediction of Solar generation of AIT Library and AIT Energy building will be trained by Neural Network in Python and simple linear regression analysis. The solar power generation highly depends on solar irradiance, air temperature, rainfall and humidity percentage. Generally, the overview methodology for solar power generation is presented in Fig 10, which starts from data collection to data analysis. The raw data were obtained from AIT Water Engineering Department and AIT Energy Department Laboratory. However, the meteorological data, to train the network is mentioned in Fig 11 which has daily data and the data were measured at the same location at AIT, therefore it does not have any differences for both buildings even though it is parted by 0.3 km, shown in Fig 9.

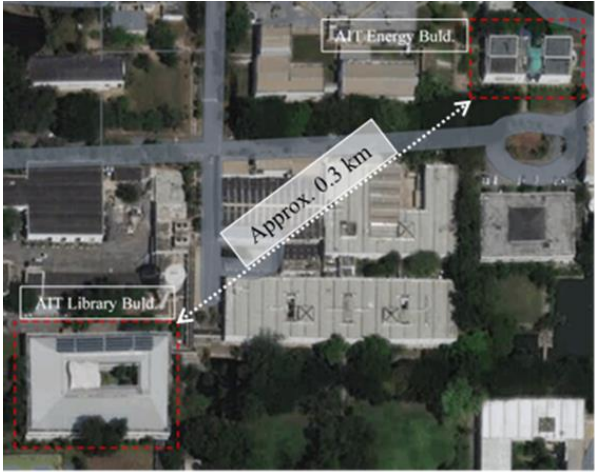

Figure 9. Location and distance between AIT Library and Energy building.

Preparation of neural network training initiates with data collection which includes raw data. The raw data has to be rearranged in order to satisfy the data processing for training. In this case, prepared datasets are separated into three scenarios, shown in Table I. After that, the neural network structure is configured from creating hidden layers with different weight ratios.

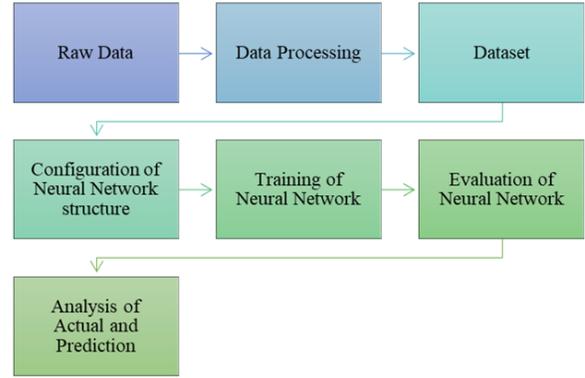

Figure 10. Overview flowchart of Neural Network training for Solar power prediction.

TABLE I. TOTAL DATASETS OF EACH TRAINING SCENARIO

| | | Solar Irradiance | Avg. Air temperature | Avg. relative humidity | Avg. rainfall |
|---|---|---|---|---|---|
| AIT ET building | ET Part.1 | 540 | 540 | 540 | 540 |
| | ET Part.2 | 61 | 61 | 61 | 61 |
| | Total | **601** | **601** | **601** | **601** |
| AIT Library building | Lib | 357 | 357 | 357 | 357 |
| | Total | **357** | **357** | **357** | **357** |

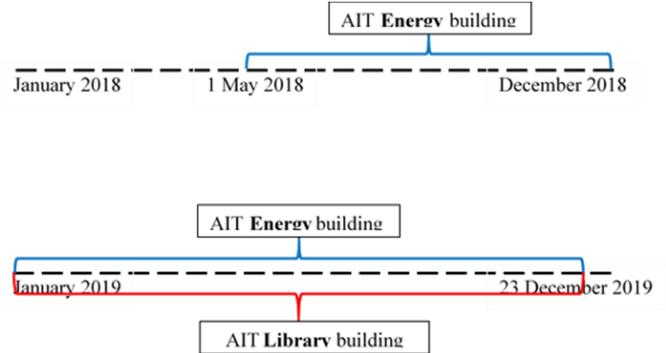

Figure 11. Dataset timeline of past solar generation of the two prosumer nodes.

The structure of neural network to train the model can be specified as Fig 12. The input parameters and hidden layers are scaled. The model includes three dense hidden layers with nodes inside of 256, 128 and 64 respectively and the weights are distributed accordingly. However, the input data will be solar irradiance, average air temperature, average relative humidity, and rainfall due to the correlation and accuracy of model training instead of using every metrological data. Adaptive learning rate method, which is Adam optimizer is used to train the model of individual learning rate for different parameters. The optimizer provides better results on accuracy than other optimizers. Datasets for every model are split into 80% to train and validation for 20% with past solar generation data, additionally, testing dataset is 100%. In here, ReLU activation function is used to train the model since it has simple computation to perform the training and the output is a continuous linear function. It identifies every positive value and makes zero to negative values.

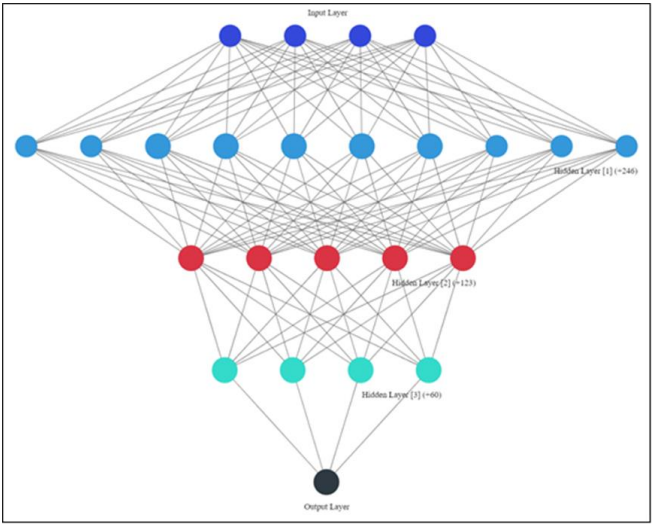

Figure 12. Layers of Neural Network of the trained model.

## III. RESULTS AND DISCUSSIONS

### A. Energy Trading at Prosumer Nodes

The simulation is performed based on the actual data of at AIT Energy and Library building. Each trading was made during the surplus hours of the solar generation at the nodes, and the tokens are transferred to the merchant's account. Mining time will be approximately forty seconds per block. The consumption is raised to 35.5 kW starts from around 7:00:00 and decreased back around 22:00:00. The solar generation data and consumption is based on the actual data of 9 October 2019. The demand request had been made randomly and the simulation time is the same as the real-time.

Forty-five transactions were created for a whole day period from 09:03:46 to 15:53:02. As for the profile of AIT Energy Building, the peak consumption is approximately 2.7 kW and the solar generation can be reached to nearly 8 kW. Therefore, the surplus is around 5 kW and the behaviours of electricity trading power from AIT Energy building to the library building is presented shown in Fig 13 with transferred power to AIT Library building and net surplus power at AIT Energy building. Simulated information of each block includes an index, timestamp, transferred token and a hash value is stated in Table IV. Automatic buying and selling are not considered in this research study, therefore, both nodes have to request and approve the transaction manually.

### B. Analyzed Prediction Data on Solar PV Generation by MLFF Neural Network

The minute ahead prediction is implemented by every five minutes intervals of power trading duration 9 October 2019. The same input parameters for every five minutes start from 00:00:00 to 23:45:00 and similar MLFF model architecture is used to train, therefore, the total steps are 286 with four input parameters. Three epochs (epoch 800, 1000, and 5000) are used to train to find the best accurate prediction result for AIT Energy building and AIT Library building.

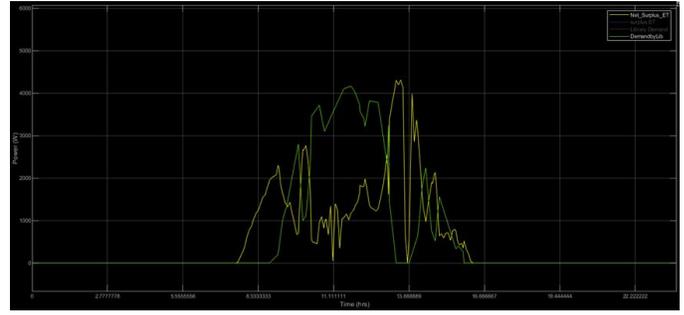

Figure 13. Demand power by AIT Library building (green) and net surplus solar energy of AIT ET building (yellow).

TABLE II. STATISTICAL RESULT OF MINUTE AHEAD PREDICTION MODEL FOR AIT LIBRARY BUILDING.

|  | AIT Library Building | | |
|---|---|---|---|
|  | Epoch 800 | Epoch 1000 | Epoch 5000 |
| Correlation <Corr>: | 0.999998842 | 0.999994862 | 0.999999614 |
| Mean Absolute Error <MAE> (Forecast): | 0.000267535 | 0.000562788 | 0.000154549 |
| Root Mean Squared Error <RMSE>: | 0.000598226 | 0.001258433 | 0.000345582 |
| Mean Square Error <MSE>: | 3.58E-07 | 1.58E-06 | 1.19E-07 |
| Mean Percentage Error <MPE>: | inf | inf | inf |
| Mean Absolute Forecast Error <MAPE>: | 0.026753489 | 0.056278817 | 0.015454888 |
| Accuracy: | 84.28% | 92.57% | 95.99% |
| Coefficient of variation <$R^2$> | 0.9781 | 0.9905 | 0.999 |

TABLE III. STATISTICAL RESULT OF MINUTE AHEAD PREDICTION MODEL FOR AIT ENERGY BUILDING.

|  | AIT Energy Building | | |
|---|---|---|---|
|  | Epoch 800 | Epoch 1000 | Epoch 5000 |
| Correlation <Corr>: | 0.999999992 | 0.999999802 | 0.999999614 |
| Mean Absolute Error <MAE> (Forecast): | 2.26E-05 | 0.000110618 | 0.000154549 |
| Root Mean Squared Error <RMSE>: | 5.05E-05 | 0.00024735 | 0.000345582 |
| Mean Square Error <MSE>: | 2.55E-09 | 6.12E-08 | 1.19E-07 |
| Mean Percentage Error <MPE>: | inf | inf | inf |
| Mean Absolute Forecast Error <MAPE>: | 0.002256706 | 0.01106184 | 0.015454888 |
| Accuracy: | 94.25% | 97.86% | 98.89% |
| Coefficient of variation <$R^2$> | 0.9971 | 0.9998 | 0.9999 |

According to Table II, the evaluations have significant values, which are nearly zero, thus training models can be accepted. Additionally, the accuracy and $R^2$ show the most accurate training scenario – epoch 5000 > epoch 1000 > epoch 800. Furthermore, based on the training result for AIT Energy building, epoch 5000 has the positive solution. The statistical result for training result of AIT Energy building is stated in Table III.

TABLE IV. TRANSACTION RECORD ON BLOCKCHAIN-BASED POWER TRADING.

| Index | Amount | Nonce | Timestamp | Hash value |
|---|---|---|---|---|
| 2 | 200 | 456 | 23-April-2020 9:03:46 | D71F0677ECCE923930E43426E60740001A21471EBBE32BC8A9E289A381EE5D86 |
| 3 | 1030 | 3469 | 23-April-2020 9:14:05 | CE315235C2BEB32E251DD4DC52619EF95E10492C77DA2C4FB549EA0DBE35EDFB |
| 4 | 1347 | 6329 | 23-April-2020 9:23:05 | FD1408378F79C6A2366CF6D535C33370839273A77E07E0FF54811AF6F73255BD |
| 5 | 2020 | 8275 | 23-April-2020 9:34:46 | E828B13793F4BDD78B729FAE932C7060DD08F2D243A9E24E841FB41C87A69625 |
| 6 | 2280 | 5279 | 23-April-2020 9:39:17 | FFAA1D3A844A5C917B678D7F9D341E7217FB48A3804B655F934B3D3AFC23C973 |
| 7 | 2800 | 2596 | 23-April-2020 9:48:56 | F4B801BD66F69ACF905A5BBD9C64EC57FCBE01F241352EF0DEF3A0E8A2461B38 |
| 8 | 1867 | 5261 | 23-April-2020 9:53:17 | 6F3ACC1A66F8E6DADCDA73133C7720F2D745752D1E2E38F92145DAC00A9F8B2E |
| 9 | 1000 | 562 | 23-April-2020 9:58:35 | C98272EC3941E5AEF498394014697C40DC7C5A1AD7C1A19E446B236E23262783 |
| 10 | 1107 | 733 | 23-April-2020 10:05:25 | A8C43F3EA544D97A37FD6EE27C67092341CCAAAB878529C536B4D8AF2123A959 |
| 11 | 2213 | 484 | 23-April-2020 10:14:35 | F2D7490157D50C86FD13C1105CE347FDBB87C01FAAC64D89787548C8809141CC |
| 12 | 3453 | 5267 | 23-April-2020 10:17:37 | 4EE831023306B02407460866DD5962D2824D6827A2BD5553174B34F44D571144 |
| 13 | 3720 | 5201 | 23-April-2020 10:34:49 | 07A3508936475AA31ABB1524E5D5066DB8F19A2C9A7DA29A88985DA34042876A |
| 14 | 3100 | 6326 | 23-April-2020 10:46:25 | 22583A87F89732157938C7964853EEF77C3C2FD1BED2CE4AB6F8071CA04BED0C |
| 15 | 3680 | 8259 | 23-April-2020 11:10:37 | 7D1A88D91C4CE01CDFA6A2F0F52AA635904C240E6ADDB31966B5799FD1F736EB |
| 16 | 4040 | 2528 | 23-April-2020 11:26:32 | 2D3C264088D90FF1232513A39447C0A6283262D7788755519A41D872D92ED3FF |
| 17 | 4100 | 5189 | 23-April-2020 11:29:01 | 97085DAB122AF554AC2439770A8A744AFCC695D35A2948F1D15889AF55AC9C4B |
| 18 | 4133 | 625 | 23-April-2020 11:36:25 | A2C4DC19DF5C88158B16E966E726D8C415FAAB930999275B7204BA8E406C4C33 |
| 19 | 4167 | 5124 | 23-April-2020 11:44:30 | 8851524BBBD8B5BA0B5C117250010DB259B60D1D2FE338498F50D47105E92A79 |
| 20 | 4047 | 1478 | 23-April-2020 11:52:53 | B4439AAD25A9DB2640BEC371CFCA859F257A8A91205E9921019A5D3FFAD2D6C3 |
| 21 | 3740 | 3520 | 23-April-2020 12:03:45 | 48CBB08EF22E5CB7BC6BD37B1E19E081A682A483571FBCD3190202F831F00E49 |
| 22 | 3567 | 255 | 23-April-2020 12:05:03 | 5FA15AAA5D830A66E35E928CBD375500C7CD6DFEB87BC1156D931FE7F63B6E44 |
| 23 | 3393 | 2450 | 23-April-2020 12:13:52 | 10DAE812B4A883E93E963B59DE0F80638EEEDD660266F20FBF3B1E82DD2D0DB7 |
| 24 | 3220 | 419 | 23-April-2020 12:16:09 | 1BE27FD2959E0303947AA7595E7103A0113AEFDBB1008ACB9042D12AA7D64D53 |
| 25 | 3820 | 252 | 23-April-2020 12:25:50 | 5C51D5F69981ADF728B96119374E79579DC037B29FC48A375EA6A03919079806 |
| 26 | 3780 | 591 | 23-April-2020 12:45:00 | C1CF75D1A4BA04844DCB44E8D3CA03242372BFAE5A122184D63C2DBE9BFCE263 |
| 27 | 2340 | 8149 | 23-April-2020 13:04:56 | C5E14485F629D867781ADF15351858BFF0618EC4C0D570ADDB99611DC3F5072D |
| 28 | 3240 | 259 | 23-April-2020 13:08:03 | E5EBFB48356E25B3CD72E457AA4F24D0E27751700A4DE2CD50B8803022633790 |
| 29 | 1440 | 629 | 23-April-2020 13:10:00 | C10DA8EEF1EAE8566F7C2B8C1E2F6EF26D70E3F86A0EAAFE4D52BB05957482BA |
| 30 | 360 | 291 | 23-April-2020 13:21:20 | 5C91E6F09596D329069B14C60C757BEE4193716DE8740E7814610F1E32153223 |
| 31 | 180 | 421 | 23-April-2020 13:25:00 | AB264041DBC7579BB69A6E12AC57EEB0EA1F13055CC4029A8B3ED19EB70B938E |
| 32 | 320 | 521 | 23-April-2020 14:04:06 | FA3937826E71F7DAB4FBC8651CB84B7EAB8C001E7DB0075C84EFDB398EB5675A |
| 33 | 640 | 251 | 23-April-2020 14:12:35 | 614029AE635D252D771F3D3B9C686A6A952DFB3F3D23E5D210B37AA39AE933AB |
| 34 | 960 | 249 | 23-April-2020 14:15:35 | 5F71E32E26AC3FE56CA164D4BC63B1F1688C5C5C4AA7AFB3271442A69B580B51 |
| 35 | 1813 | 4519 | 23-April-2020 14:22:25 | 5B14F6289A3BC2B027B379903E47E1175E2B906204D0D2A8F30A1FCA1A9C05E7 |
| 36 | 2240 | 259 | 23-April-2020 14:30:26 | CF45B0654BBC65E65911BBB203942B9E85C56EE4E105B93DA95B983C9EE42D52 |
| 37 | 1493 | 529 | 23-April-2020 14:36:37 | 8A2A2E9749F0E52D22D64078755F9EEAA4409C0E91B0443F4F8A06C78743126C |
| 38 | 747 | 6791 | 23-April-2020 14:43:20 | F43612E82BAF8B364B22D15943A91BE232E480D7B21B1D20246919A5C657427D |
| 39 | 520 | 1588 | 23-April-2020 14:51:25 | 68D35BA3CCBBD2D4D21E942F21D49B5E721831B9024A4AFB1B4503E67022643C |
| 40 | 1040 | 819 | 23-April-2020 14:55:25 | 907C987CE796D1C3C76EDCDB8ABC2DE634740DFC6E55903CD73B057930CE9F91 |
| 41 | 1560 | 2501 | 23-April-2020 15:00:27 | 8EA63654F87BCAE90BFBB47C6F7EAAF1AEBF831A2AAE2E839342FC62D67FDFEA |
| 42 | 300 | 5196 | 23-April-2020 15:31:25 | 580AA26E7D6AC3EAF9C2BEA415733C48C7FD216F55904A4A9F709C0E33863565 |
| 43 | 333 | 9158 | 23-April-2020 15:36:56 | 8ACA9E5D998CD54CF7BE753E80870465428C6DD71A681D5D8889880FD881C5AB |
| 44 | 400 | 5920 | 23-April-2020 15:42:25 | 3CD5CB63CA589C2282F69C7E1AEC2F593DBACA82263A968978A2B9692FC20673 |
| 45 | 267 | 6156 | 23-April-2020 15:53:02 | B0319683F9793F6E95E7962B21BA2542734BDDFF0028705CB0F3CAFFEDAB122C |
| 46 | 133 | 1632 | 23-April-2020 15:53:53 | 3C81F2561231A02900ECA0D2509892C47B51E22EC924B9F7EAAD352CE3F3DE47 |

TABLE V. EVALUATION OF EFFECTIVE MODELS IN THREE DIFFERENT EPOCHS.

| | AIT Energy building | | | | | | AIT Library building | | |
|---|---|---|---|---|---|---|---|---|---|
| | Epochs 800 | | Epochs 1000 | | Epochs 5000 | | Epochs 800 | Epochs 1000 | Epochs 5000 |
| Mean square error | 0.044% | 0.30% | 0.00668% | 0.000866% | 0.0304% | 0.361% | 0.128% | 0.000139% | 0.383% |
| Mean absolute forecast error | 0.94% | 2.47% | 0.365% | 1.316% | 0.780% | 2.689% | 1.60% | 5.27% | 0.171% |
| Root mean squared error | 2.11% | 5.52% | 0.817% | 2.94% | 1.744% | 6.013% | 3.58% | 0.817% | 0.00147% |
| Accuracy | 92.43% | | 92.83% | | 93.80% | | 95.53% | 90.99% | 93.62% |
| Coefficient of variation ($R^2$) | 90.12% | | 94.80% | | 95.97% | | 73.78% | 53.94% | 66.21% |

Based on the statistical result, the most accurate training results are selected to describe the comparison of actual data and predicted data for both AIT Library and Energy building. Therefore, epoch 5000 is used that has most accurate result and shown in Fig 14 and Fig 15 respectively for both buildings. Output time range is between 0:00:00 to 23:45:00, however, the visualization presents the time frame of solar generation duration. During the period of power trading, the practice of minute ahead prediction can be utilized, therefore not only seller but also buyer can straightforwardly know the upcoming generation. The output of prediction value sets by training on Neural Network for AIT Energy building is from May 1, 2018 to December 23, 2019 and for AIT Library building is from January 1, 2019 to December 23, 2019. Comparison of observed values and predicted values for AIT Energy building with three scenarios.

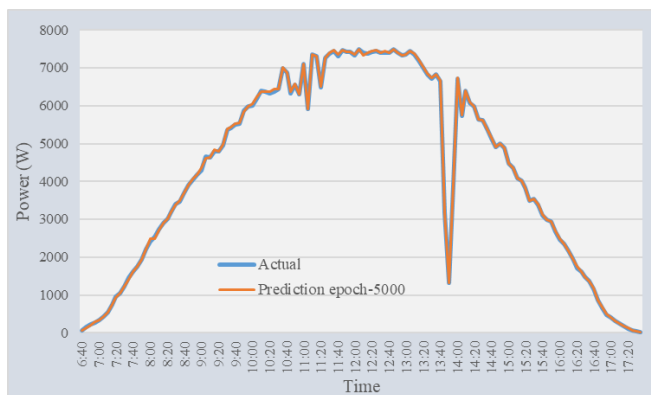

Figure 14. Comparison of actual data and minute ahead prediction of one-day solar generation at AIT Energy building.

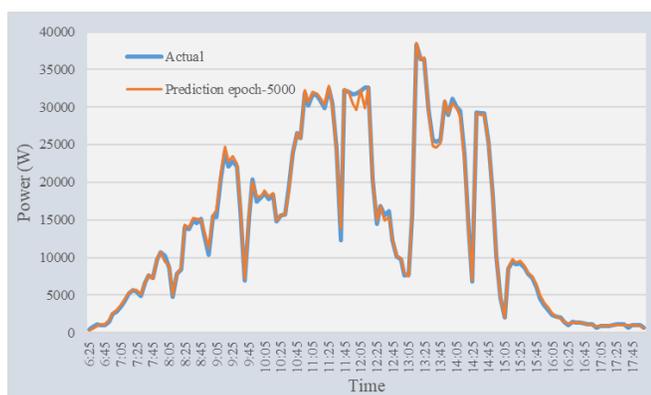

Figure 15. Comparison of actual data and minute ahead prediction of solar generation at AIT Library building during trading day.

Consequently, the loss and accuracy result of both nodes, which are the energy building and the library building, in three different epochs are stated in Table V, by means of MSE, MAFE, RMSE, accuracy, and coefficient of variation ($R^2$). In the case of AIT Energy building, from the point of view of accuracy that is between actual data and predicted data, consequently, 93.80% > 92.83% > 92.43% which means epoch 5000 > 1000 > 800. Coefficient of variation shows the same result - 95.97% > 94.8% > 90.12%, which means prediction values at epoch 5000 has the highest accuracy than epoch 1000 and 800. The accuracy of the training result of AIT library building can be stated as 95.53% > 93.62% > 90.99% and for $R^2$ - 73.78% > 66.21% > 66.21%. Therefore, one of the best epochs is 800 with the highest accuracy value than the others.

IV. CONCLUSIONS

This research shows the simulation of simple blockchain based prosumer consortium model without injection of any cryptocurrency and web service in the blockchain system. This market simulation can be based on the generation of the equation of solar PV generation, however, observed data were used to imitate to simulate in this study. Virtual tokens are stored in text file to create simple token transfer and it will use directory system, moreover, the hash environment utilizes by Cryptography SHA-256. Forty-five discrete trading were reached during the surplus hours from AIT Energy building to Library building. Parallel computing was used to run the calculating and it took significant seconds to establish create a block as well. This paper provides the future ahead prediction such as five minutes ahead and one day ahead prediction of solar generation at the prosumer nodes. Multi-layer Feed Forward neural network offers the most accurate and effective output *y* for the prediction which might consequence the behaviour of the buyers and sellers. Based on the past generation data and meteorological data, MAFE and MAPE of the models indicate the smallest error percentage with higher correlation, and the simple linear regression verifies the accuracy of the models in term of the performance selection. As for the suggestion for the future work, forecasting of load consumption of both buildings based on the geographic location, cost, capacity sizing and past consumption can be achieved based on the same model in order to understand the balance between required power and surplus power. This study can be one of the movements of paradigm shift for power trading market model based on a decentralized blockchain system, in a community and a university grid system.

ACKNOWLEDGMENT

The author gratefully acknowledges the support by Prof. Weerakorn Ongsakul, Dr. Jai Govind Singh and Dr. Chutiporn Anutrariya for their supervision during the research study, in addition my kindness toward Dr. Hien Vu Duc from AIT Energy laboratory for the data and required materials. Fast and foremost, the author would like to thank to Dr. Nimal Madhu Manjiparambil for his kind support during the shaping for this paper.